\documentclass[11pt,oneside,notitlepage,abstracton,a4paper]{scrartcl}
\usepackage{epsfig,scrpage2,graphicx}

\renewcommand{\headfont}{\normalfont}

\begin{document}

\title{\vspace*{-1.8cm}
\begin{flushright}
{\bf\normalsize LAL/RT 06-05}\\
\vspace*{-0.4cm}
{\normalsize\bf EUROTeV-Report-2005-026}\\
\vspace*{-0.4cm}
{\normalsize June 2006}
\end{flushright}
\vspace*{2,5cm}
{\LARGE Particle tracking in the ILC extraction 
lines\\ with DIMAD and BDSIM}}

\author{{\bf\large R. Appleby,}\\ 
{\it\normalsize The Cockcroft Institute and the University of Manchester,}\\ 
{\it\normalsize Oxford Road, Manchester, M13 9PL, England
\vspace*{0,3cm}}\\
{\bf\large P. Bambade, O. Dadoun,}\\ 
{\it\normalsize Laboratoire de l'Acc\'el\'erateur Lin\'eaire,}\\ 
{\it\normalsize IN2P3-CNRS et Universit\'e de Paris-Sud 11, BP 34,}\\ 
{\it\normalsize 91898 Orsay cedex, France\vspace*{0,3cm}}\\
{\bf\large A. Ferrari},\\
{\it\normalsize Department of Nuclear and Particle Physics,}\\
{\it\normalsize  Box 535, Uppsala University, 751 21 Uppsala, Sweden}}


\maketitle
\vspace*{0,5cm}
\begin{abstract}

\noindent
The study of beam transport is of central importance to the design and
performance assessment of modern particle accelerators. In this paper, 
we benchmark two contemporary codes, DIMAD and BDSIM, the latter being a 
relatively new tracking code built within the framework of GEANT4. We 
consider both the 20~mrad and 2~mrad extraction lines of the 500~GeV 
International Linear Collider (ILC) and we perform particle tracking 
studies of heavily disrupted post-collision electron beams. We find 
that the two codes give an almost equivalent description of the beam 
transport.
\end{abstract}

\newpage

\section{Introduction}

In a $e^+e^-$ linear collider such as ILC~\cite{ilc}, the beams must be 
focused to extremely small spot sizes in order to achieve high charge 
densities and, in turn, to reach the desired luminosity. Because of the 
extremely small transverse dimensions of the colliding beams, electrons 
and positrons experience very strong transverse electromagnetic fields 
at the interaction point, which leads to the emission of beamstrahlung 
photons, as well as large angular divergence and energy spread for the 
disrupted beams. A careful design of the extraction lines must therefore 
be performed to transport the outgoing beams and the beamstrahlung 
photons from the interaction point to their dumps with small losses. 
At ILC, two configurations are being studied for the crossing angle 
at the interaction point and, as a result, for the design of the 
post-collision lines. With a 2~mrad crossing angle, the main challenge 
is the extraction of the disrupted beam, which has to be achieved by 
sending the outgoing beam off-center in the first large super-conducting 
defocusing quadrupole of the final focus beam line, as well as in the two 
nearby sextupoles~\cite{Appleby:2004df,Appleby:2005nh}. On the other hand, 
with a 20~mrad crossing angle~\cite{ilc20a}, one must deal with technical 
difficulties such as large crab-crossing corrections or the construction 
of compact super-conducting quadrupoles for the incoming beam lines, as 
well as with the passage of the beams through the solenoid field with 
an angle. For the design of the ILC extraction lines, it is 
essential to have a reliable simulation program for particle tracking. 
In this study, we present a comparison between two codes, DIMAD~\cite{dimad} 
and BDSIM~\cite{bdsim}, using the present versions of the ILC post-collision 
lines for benchmarking purposes, in both large and small crossing angle 
cases.\\

The DIMAD program specifically aims at studying the behaviour of
particles in circular machines or beam lines, by computing their
trajectories using the second order matrix formalism~\cite{matrix}. 
The present version of the code makes sure that the matrix treatment 
remains correct to all orders for energy deviations~\cite{dimad}. This is 
important here, as the ILC disrupted beams downstream of the interaction 
point can have very large energy spreads. The BDSIM program~\cite{bdsim} 
uses the closed solutions in linear elements, whilst for higher-order 
elements, a GEANT4-like stepper integration method is used. The program 
is written in GEANT4~\cite{geant4} and provides a toolkit to fully simulate
interactions of particles with the surrounding matter once they have left
the vacuum pipe. However, for the purpose of this study, we only aim at 
comparing the tracking procedures in DIMAD and BDSIM: a more detailed 
evaluation of the losses, with all particle-matter interactions and with 
the subsequent background generation, is underway and will be the subject 
of a future report. In order to compare the tracking procedures in DIMAD 
and BDSIM, we will proceed as follows. In Section~2, we consider the ILC 
extraction line with a 20~mrad crossing angle, where the disrupted beam 
remains centered in all magnetic elements. We compare single particle 
trajectories as well as beam transverse spectra, as they are obtained 
with DIMAD and BDSIM at various positions along the extraction line. 
Then, in Section~3, we perform a similar analysis with the ILC 2~mrad 
post-collision line, where the geometry is slightly more complicated, 
since the disrupted beam goes off-center in the first magnetic elements. 
Finally, a summary is given in Section~4. 

\section{DIMAD and BDSIM tracking in the 20~mrad extraction line}

\begin{figure}[h!]   
\begin{center} 
\epsfig{file=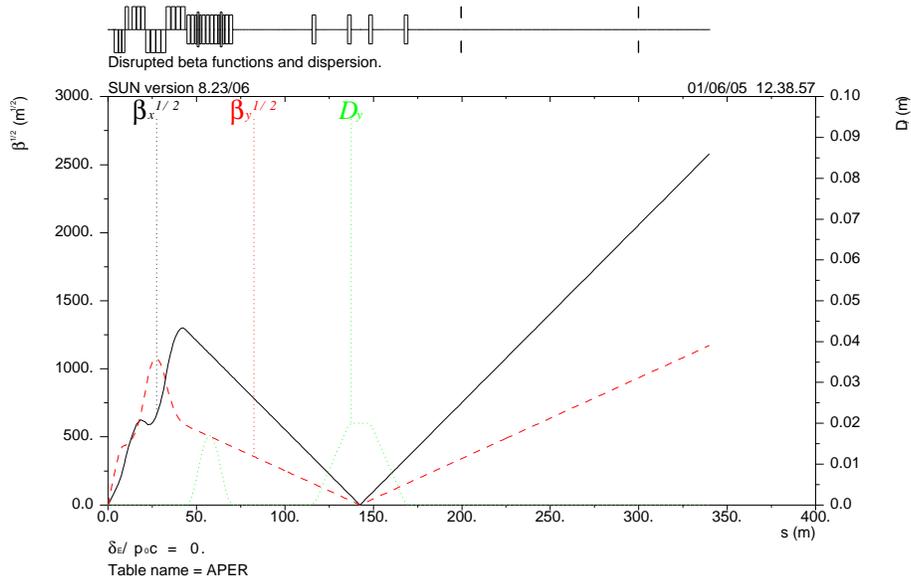,angle=-90,width=12cm}
\caption[]
{\it Betatron functions and vertical dispersion along the ILC extraction 
line with a 20~mrad crossing angle (this is an update of the lattice 
described in~\cite{ilc20a}).}
\label{ilc20optics}
\end{center}
\vspace*{-0,5cm}
\end{figure}

In order to compare the tracking procedures in DIMAD and BDSIM, we 
first consider the ILC 20~mrad extraction line. Thanks to the large 
crossing angle, one can use a dedicated line to transport each outgoing  
beam from the interaction point to its dump. In the present design 
of the ILC 20~mrad extraction line~\cite{ilc20a}, the disrupted beam 
and the beamstrahlung photons go through the same magnets to a common 
dump. The optics consists of a DFDF quadruplet, followed by two 
vertical chicanes for energy and polarization measurements and a 
long field-free region that allows the beam to grow naturally, with 
two round collimators to reduce the maximum beam size at the dump. 
Figure~\ref{ilc20optics} shows the betatron functions and the vertical 
dispersion in this design. 

\subsection{Single off-momentum particles}

\begin{figure}[h]  
\begin{center} 
\epsfig{file=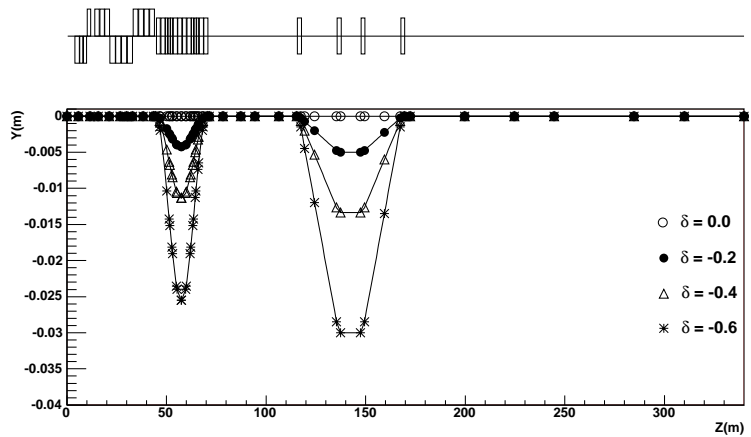,height=6.cm}
\caption[]
{\it Particle trajectories along the ILC 20~mrad extraction line, as 
computed by DIMAD (lines) and BDSIM (points), for various energy 
spreads. All particles are generated at the interaction point with 
$x=0$, $x'=0$, $y=0$, $y'=0$.}
\label{single20}
\end{center}
\end{figure}

For the sake of simplicity, we switch off any type of particle-matter 
interaction, including for the moment synchrotron radiation, in BDSIM, 
since we want to benchmark the tracking procedures only. Let us first 
compare single particle trajectories. For this purpose, we track four 
particles with ideal transverse coordinates ($x=0$, $x'=0$, $y=0$, $y'=0$) 
at the interaction point and increasing fractional energy deviation $\delta$. 
The first one has the nominal energy ($\delta$=0) and it thus follows a 
centered reference path in all elements of the extraction line. The three 
other particles have lower energies ($\delta < 0$): as a result, they 
follow different paths inside the magnetic chicanes, as shown in 
Figure~\ref{single20}. Note however that, since the total vertical 
dispersion of both chicanes is equal to zero, all particles remain on 
the same trajectory downstream of the chicanes. For all energies, there 
is a perfect agreement between the single particle trajectories obtained 
with DIMAD (with no synchrotron radiation) and with the tracking procedure 
of BDSIM.

\begin{figure}[h!]   
\begin{center} 
\epsfig{file=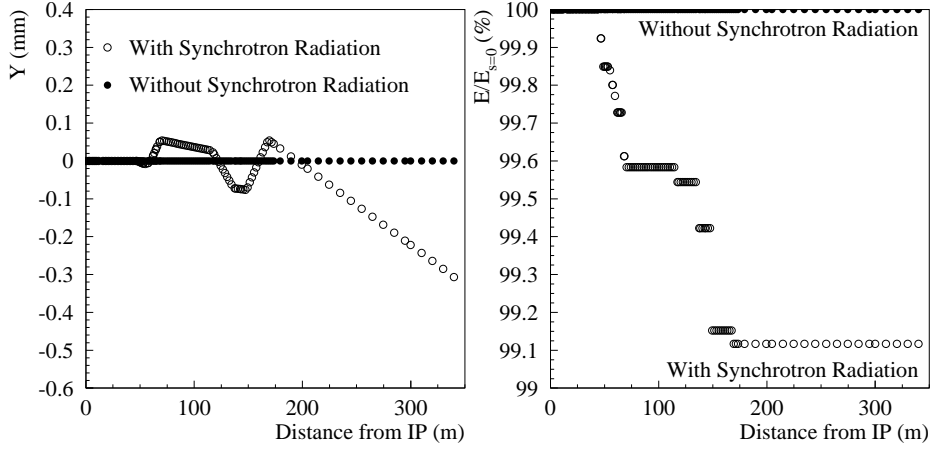,width=13cm}
\caption[]
{\it Effect of the synchrotron radiation on the tracking in DIMAD for a 
500~GeV electron ($\delta=0$).}
\label{sr00}
\end{center}
\end{figure}

\begin{figure}[h!] 
\vspace*{-0,5cm}  
\begin{center} 
\epsfig{file=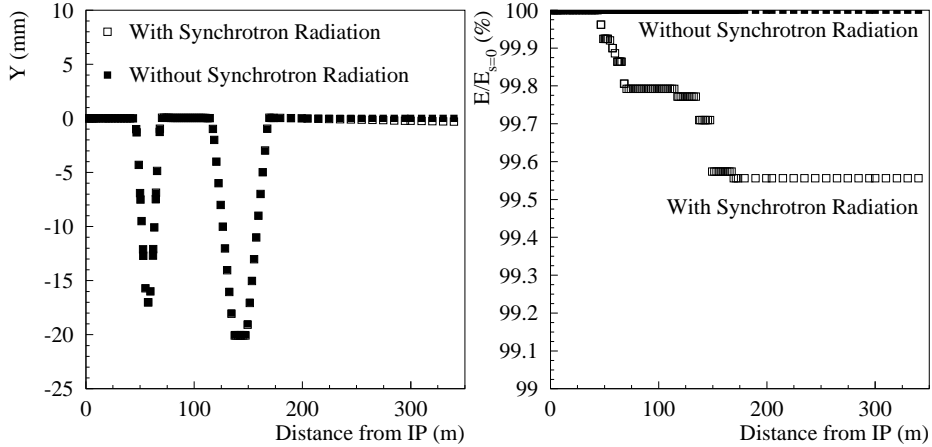,width=13cm}
\caption[]
{\it Effect of the synchrotron radiation on the tracking in DIMAD for an 
electron with $\delta=-0.5$ (E = 250~GeV).}
\label{sr50}
\end{center}
\end{figure}

Note that the paths are slightly different when the synchrotron radiation is 
taken into account. To illustrate this, let us track one 500~GeV electron 
along the ILC 20~mrad post-collision line with DIMAD, with and without 
synchrotron radiation, for $\delta=0$ (see Figure~\ref{sr00}) or 
$\delta=-0.5$ (see Figure~\ref{sr50}).\\

The energy loss due to synchrotron radiation along the ILC post-collision 
line is larger for an electron with $\delta=0$ than for an electron with 
$\delta=-0.5$, as expected. When the electron radiates a fraction of 
its energy, it leaves the reference path inside the bending magnets. 
Figures~\ref{sr00} and~\ref{sr50} show that this effect is relatively 
more important for particles having the nominal energy (which 
should be centered in all magnetic elements) than for particles 
with a non-zero energy deviation (which are already passing off-center 
in all magnetic elements). Note that the losses occuring during the 
transport of the disrupted beams from the interaction point to the 
dumps concern almost exclusively particles with a large energy deviation 
and almost never particles close to the nominal energy. Therefore, we expect 
the synchrotron radiation to have a limited influence on the power lost 
along the ILC extraction line and, for the rest of this study, we will 
switch off this effect in both DIMAD and BDSIM.

\subsection{Complete phase-space}

\begin{figure}[h!]   
\begin{center} 
\epsfig{file=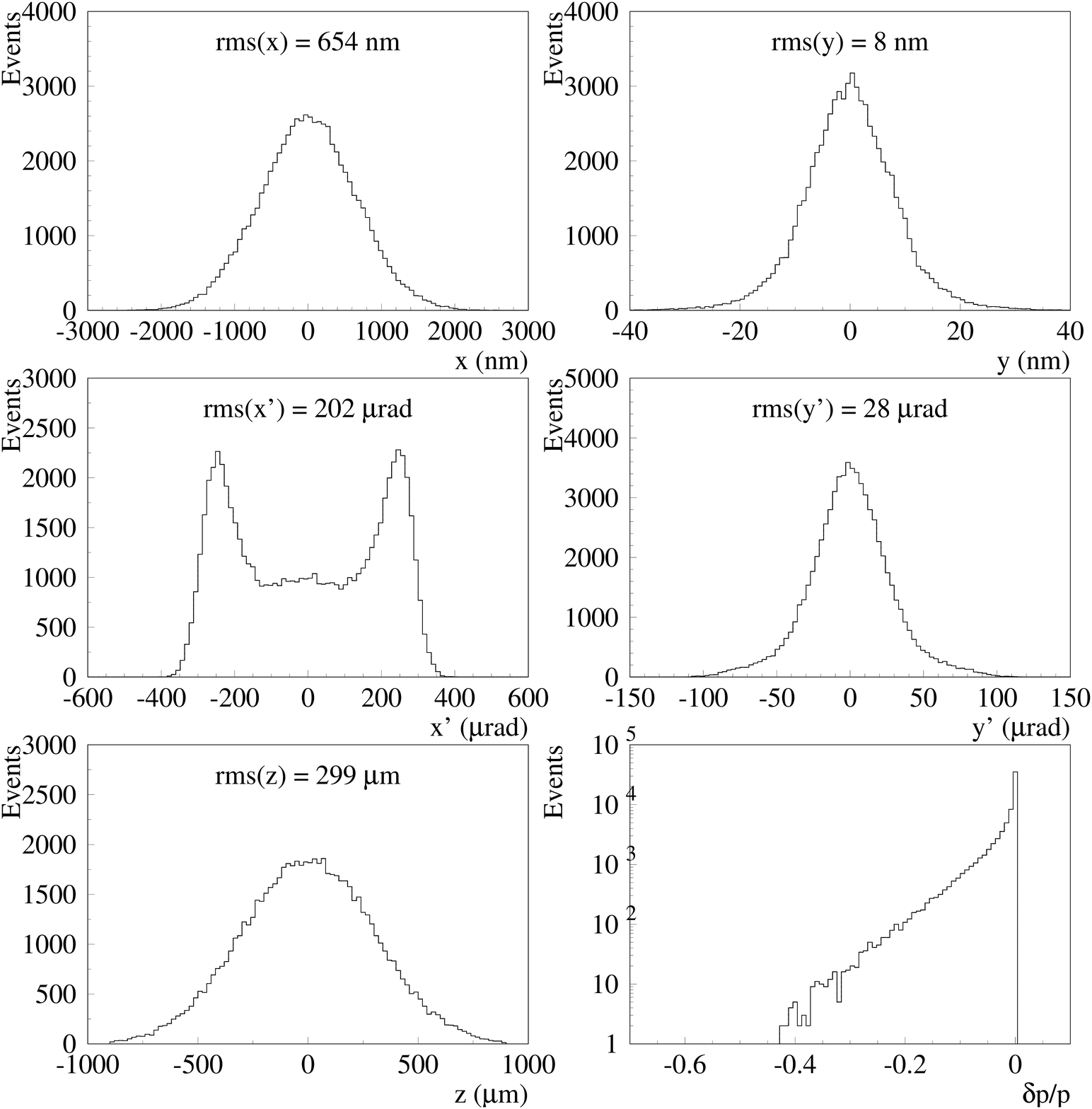,width=12cm}
\caption[]
{\it Transverse and longitudinal distributions of the disrupted beams at 
the ILC interaction point for a centre-of-mass energy of 500~GeV in 
the nominal luminosity configuration. Here, about 70000 $e^+$/$e^-$ 
macro-particles are displayed.}
\label{ilc500}
\end{center}
\end{figure}

Having shown that DIMAD and BDSIM agree perfectly when following single 
particles with various energy deviations, let us now compare transverse 
distributions of particles at several locations along the ILC extraction 
line. For this purpose, since we are not interested in a detailed estimation 
of the losses along the post-collision line but only in the behaviour of the 
tracking in both simulations, we use the suggested nominal beam parameters 
for ILC at a centre-of-mass energy of 500~GeV~\cite{sugg}, for 
which the beam transport from the interaction point to the dump 
is almost loss-free (at least with the low-statistics input files 
that we use for this study). The corresponding particle distributions 
for the $e^+/e^-$ disrupted beams at the interaction point, just after 
the bunch crossing, obtained with the GUINEA-PIG program~\cite{guinea-pig} 
are shown in Figure~\ref{ilc500}.

\begin{figure}[h!]   
\begin{center} 
\epsfig{file=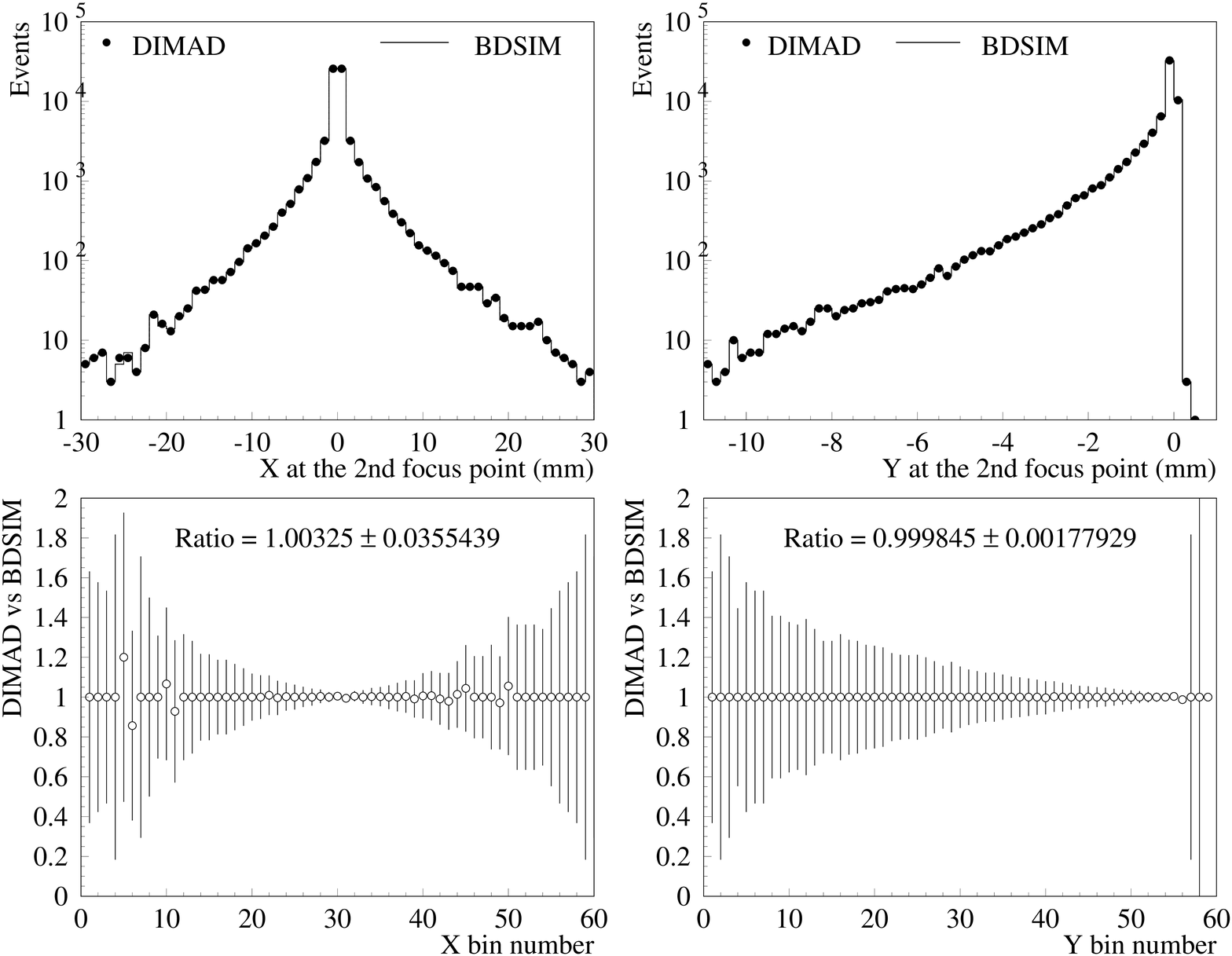,height=8.5cm}
\caption[]
{\it Comparison of the transverse beam distributions obtained with DIMAD 
(full circles) and BDSIM (line histogram) at the secondary focus point 
MEXFOC. Both upper plots are distributed over 60 bins. The bottom plots 
show the ratio of the DIMAD and BDSIM distributions (see text for details).}
\label{focilc20}
\vspace*{5mm}
\epsfig{file=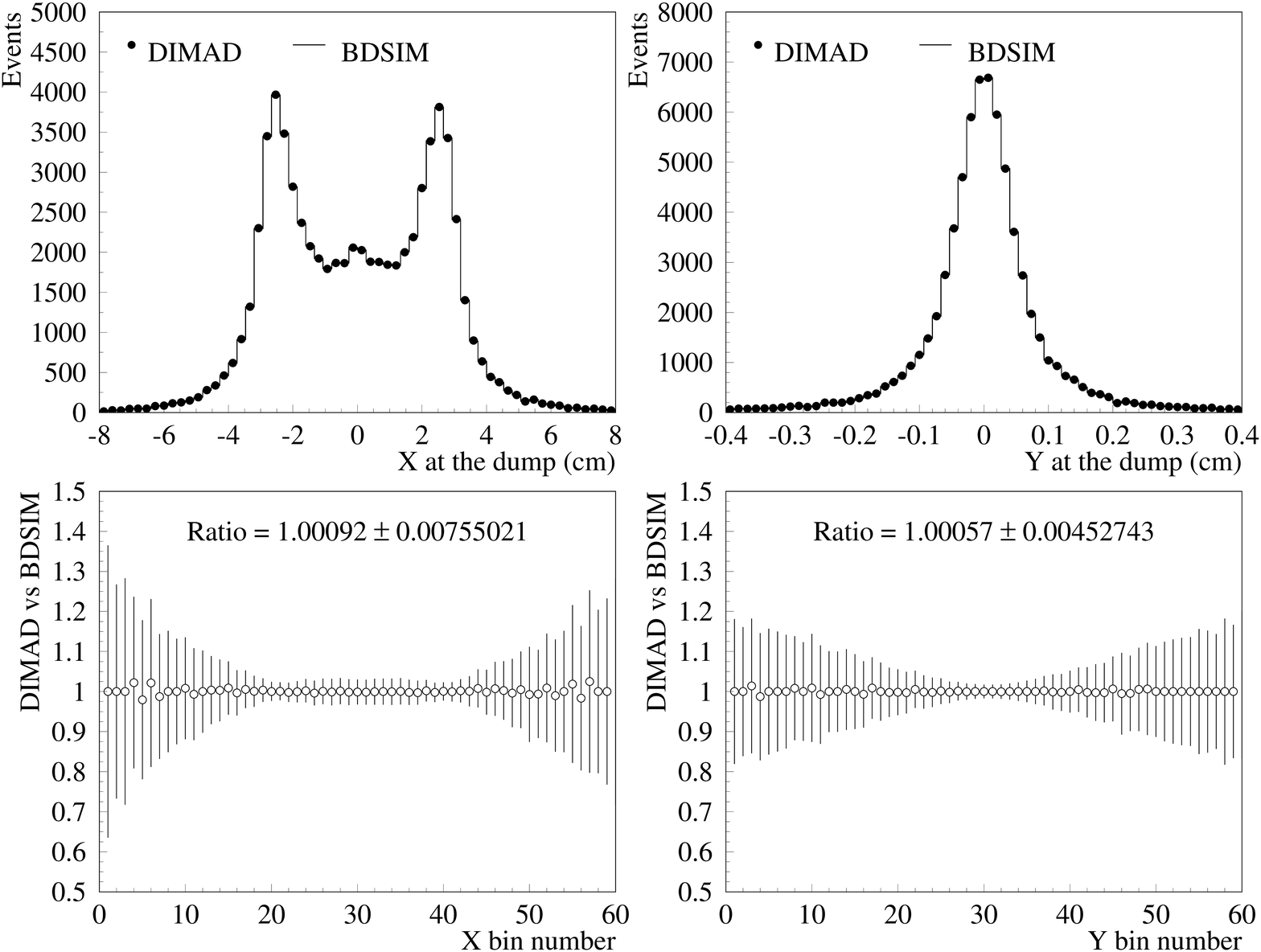,height=8.5cm}
\caption[]
{\it Same as Figure~\ref{focilc20}, obtained at the end of the 20~mrad 
extraction line.}
\label{dumpilc20}
\end{center}
\end{figure}

Such particle distributions are transported along the ILC 20~mrad 
extraction line with either DIMAD or BDSIM. At several locations 
of interest (typically before, in and after each vertical magnetic 
chicane), we project the transverse beam distributions obtained with 
each program into binned histograms and we compare them quantitatively. 
An illustration of this procedure is shown in Figures~\ref{focilc20} 
and~\ref{dumpilc20} for the transverse beam distributions ($x$ and $y$) 
that are obtained respectively at the secondary focus point MEXFOC, located 
at $s = 142.4~\mbox{m}$ (where $\beta_x$ and $\beta_y$ are very small, 
with a vertical dispersion of 2~cm) and at the dump. The open 
circles show the ratio between the number of events found by DIMAD 
or BDSIM in a given histogram bin and the error bars account for the 
limited number of events per bin (very few events are found in the 
tails, which explains the large error bars there).

The transverse distributions of the disrupted beams were also computed with 
DIMAD and BDSIM at several other locations in the 20~mrad extraction line. 
Their comparison also showed excellent agreement.

\section{DIMAD and BDSIM tracking in the 2~mrad extraction line}

\begin{figure}[h]
\begin{center}
\includegraphics[width=9cm,angle=-90]{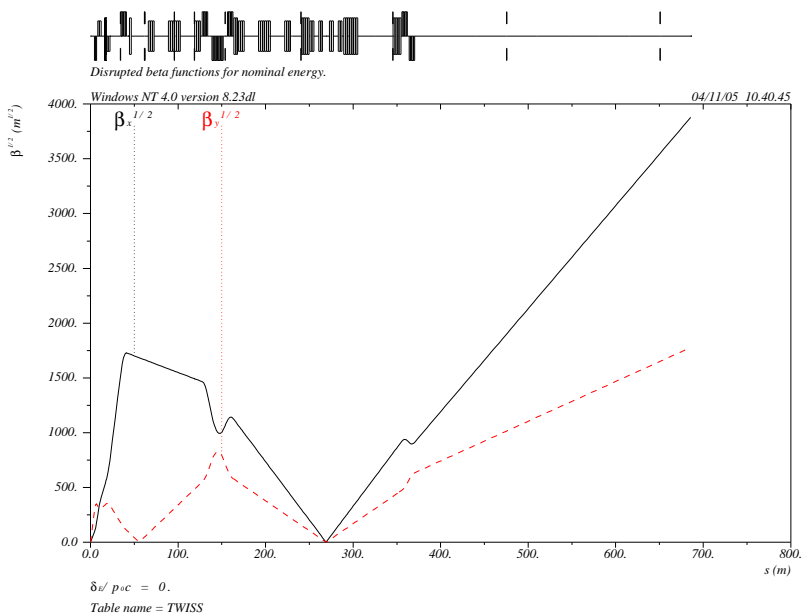}
\vspace*{-0,5cm}
\caption{\it Betatron functions along the ILC 2~mrad post-collision line.}
\label{fig2mradoptics1}
\end{center}
\vspace*{-0,5cm}
\end{figure}

\begin{figure}[h]
\begin{center}
\includegraphics[width=10cm,angle=-90]{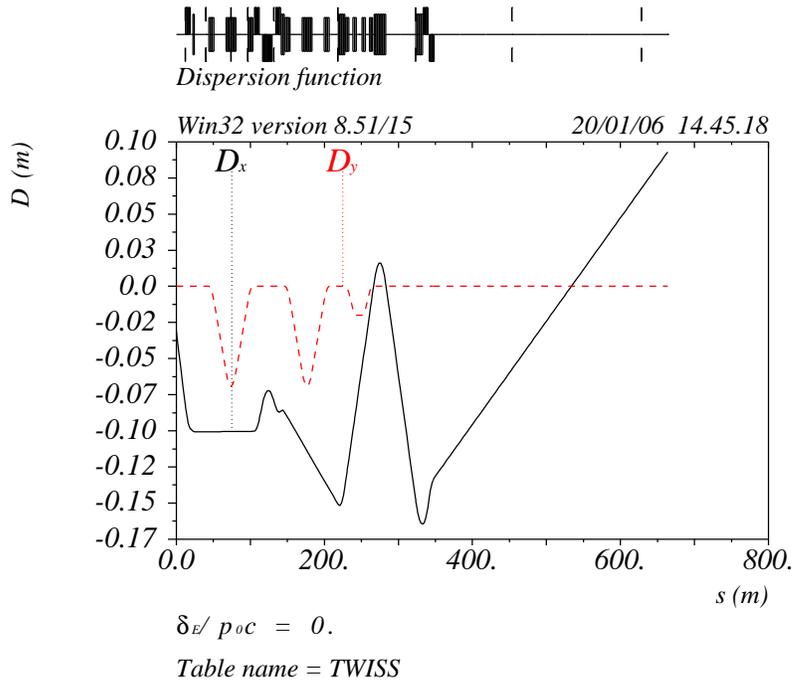}
\vspace*{-0,5cm}
\caption{\it Dispersion functions along the ILC 2~mrad post-collision line, 
downstream of the final doublet of the incoming beam line.}
\label{fig2mradoptics2}
\end{center}
\vspace*{-0,5cm}
\end{figure}

In this section, we shall compare the DIMAD and BDSIM tracking procedures 
for the ILC 2~mrad extraction line~\cite{Appleby:2004df,Appleby:2005nh}. 
This layout was developed in an attempt to preserve the physics advantages of 
the head-on scheme suggested in the TESLA TDR~\cite{tesla}, whilst mitigating 
the associated technological challenges. In this scheme, the colliding beams 
cross with a small horizontal angle of around 2~mrad. The outgoing disrupted 
beam then passes off-axis through the first magnets of the incoming final 
focus beam line, so these magnets require a large magnetic bore. In the 
design used for this work, the outgoing beam passes through the bore of 
the final quadrupole QD0, both final sextupoles but not the second-to-final 
quadrupole QF1. The outgoing beam sees however the pocket field of this 
latter magnet. The strongest design challenge lies in this shared doublet 
region, with current work focusing on the choice of final doublet magnet 
technology~\cite{Appleby:2005nh}. The extraction line of the 2~mrad scheme 
follows the 20~mrad design, with the inclusion of downstream diagnostic 
structures. The current version of the optics was presented at Snowmass 
2005~\cite{yuri,snowmass05}. The corresponding linear optics is shown in 
Figures~\ref{fig2mradoptics1} and~\ref{fig2mradoptics2}.

Following the doublet, the beam is focused with a series of large bore 
quadrupoles. This is followed by an energy clean-up vertical chicane, 
diagnostic chicanes for the purpose of energy spectrometry and polarimetry 
and, in the same way as in the 20~mrad scheme, a long field-free region to 
allow the beam to grow to the dump. Note that, in the 2~mrad scheme, separate 
dumps are used for the charged beam and for the beamstrahlung photons.

\subsection{Single off-momentum particles}

We shall now, in exactly the same way as for the 20~mrad case, track 
single particles. We consider three particles at the interaction point, 
with ideal transverse coordinates and energy deviations $\delta$=0, 
-0.2 and -0.4. The first particle traces out the nominal trajectory. 
The off-momemtum particles trace out different trajectories, just as 
in the 20~mrad case, and the same comments with regard to the downstream 
chicanes apply. For all momentum deviations, we see a perfect agreement 
between DIMAD and BDSIM.\\

\begin{figure}[h!]   
\begin{center} 
\epsfig{file=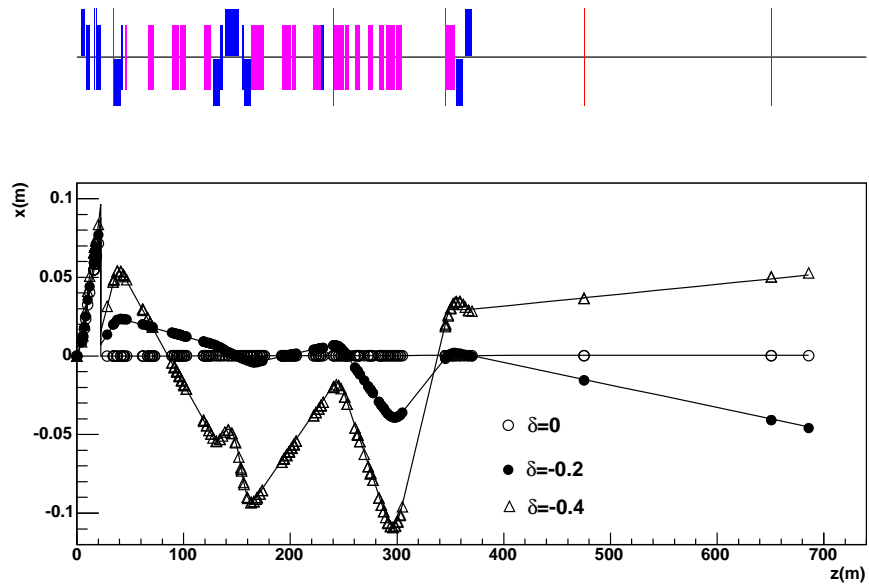,width=12cm}
\caption[]
{\it Particle trajectories along the ILC 2~mrad extraction line, as computed
with DIMAD (lines) and BDSIM (points), for various energy spreads. All
particles are generated at the interaction point with $x=0$, $x'=0$, $y=0$, 
$y'=0$.}
\label{figsec31}
\end{center}
\end{figure}

The 2~mrad case has an extra degree of complexity compared to the
20~mrad case. This is because the beam is off-axis in the final
doublet, including in one case the pocket field, which requires the 
introduction of a multipole expansion in BDSIM. The extraction line 
after the final doublet is then aligned to ensure that the outgoing 
beam is on-axis in this part of the beamline. For the single particle
tracking, we align the extraction line to the nominal particle. When
we consider beam distributions (for example in the next section), we 
align the extraction line after the final doublet to the outgoing beam 
centroid. This transformation is implemented as an active transformation
of the beam in DIMAD and as a 3D transformation of the reference coordinate 
system in BDSIM. Note that the shift of the particle trajectories can be 
seen in Figure~\ref{figsec31}.

\subsection{Complete phase-space}

\begin{figure}[h!]
\begin{center} 
\epsfig{file=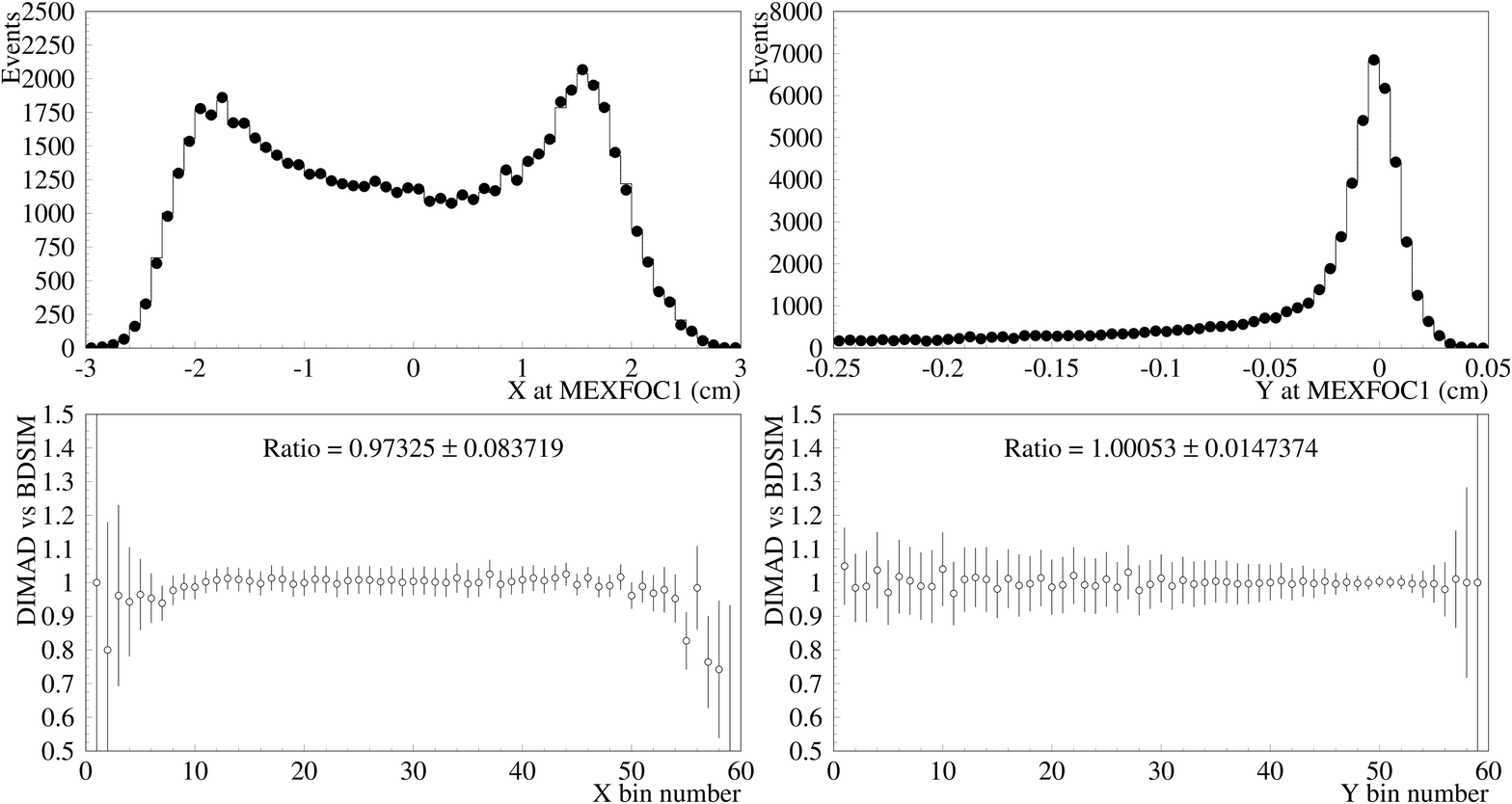,height=8.5cm}
\caption[]
{\it Comparison of the transverse beam distributions obtained with DIMAD 
(full circles) and BDSIM (line histogram) at MEXFOC1. Both upper plots 
are distributed over 60 bins. The bottom plots show the ratio of the 
DIMAD and BDSIM distributions.}
\label{mexfoc1}
\end{center}
\end{figure}

\begin{figure}[h!]
\begin{center} 
\epsfig{file=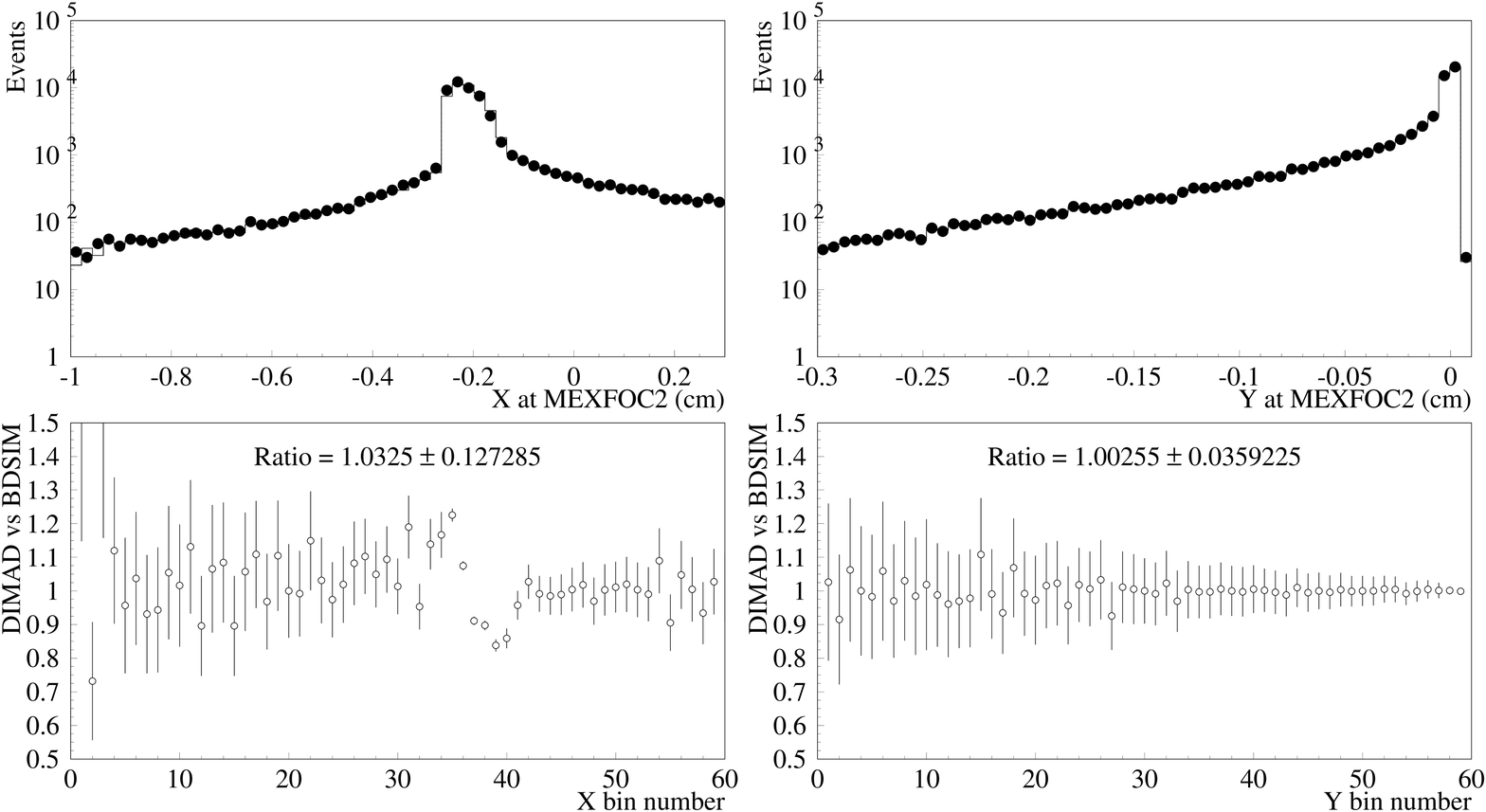,height=8.3cm}
\caption[]
{\it Same as Figure~\ref{mexfoc1}, obtained at MEXFOC2.}
\label{mexfoc2}
\vspace*{3mm}
\epsfig{file=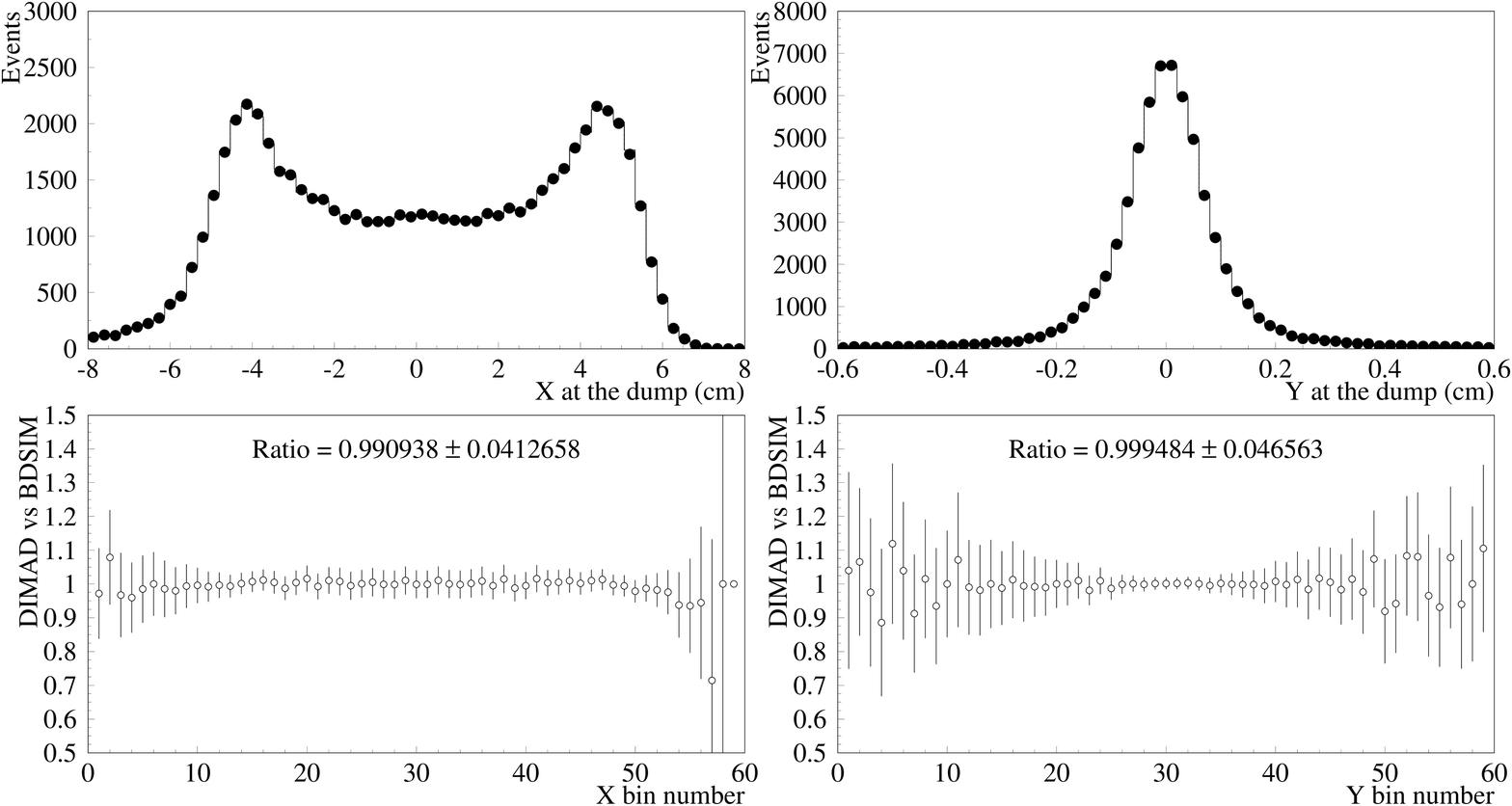,height=8.cm}
\caption[]
{\it Same as Figure~\ref{mexfoc1}, obtained at the dump.}
\label{dump2}
\end{center}
\end{figure}

Following the discussion of the 20~mrad BDSIM/DIMAD comparison in the 
previous section, we now simply describe the results of the comparison 
for the 2~mrad case. The studies in this section were performed for a 
250~GeV disrupted beam with the same nominal parameters as in the previous 
section (see Figure~\ref{ilc500}). 
Figures~\ref{mexfoc1} to~\ref{dump2} show the results of the comparison at 
three extraction line locations. Figures~\ref{mexfoc1} and~\ref{mexfoc2} 
show the comparison at MEXFOC1, which is located after the energy 
clean-up chicane, and at MEXFOC2, which is the secondary focus of the 
polarimetry chicane, respectively. As for Figure~\ref{dump2}, it shows 
the comparison at the beam dump.\\

In these plots, chosen to be at various places of interest, we have projected 
the transverse beam distibutions obtained from the tracking into bins, and 
we have formed the ratio of the DIMAD prediction to the BDSIM prediction. As 
in the previous section, the open circles show the ratio, with the error bars 
accounting for the limited number of events in a given bin (again, the larger 
error bars are from the beam tail where there are less events). All diagrams 
show a good agreement between DIMAD and BDSIM for the ILC 2~mrad extraction 
line, except at the secondary focus of the polarimetry chicane (MEXFOC2), 
where some slight discrepancies are visible. These may be due to the 
different treatments of high-order effects in the optical transport for 
non-linear elements (see Section~1).

\section{Conclusion}

In this paper, we performed a detailed benchmarking study of two particle 
tracking codes, DIMAD and BDSIM. For this purpose, we have considered 
the ILC extraction lines with a crossing angle of 2~mrad or 20~mrad 
and, in each of these two configurations, we have performed tracking 
studies of heavily disrupted post-collision electron beams. Here, only 
the nominal luminosity case of the 500~GeV machine was studied. We find 
that both programs give an equivalent description of the beam transport 
in all parts of the post-collision lines, except at the secondary focus 
for the 2~mrad design.\\
 
A similar benchmarking study is presently being performed in order to 
compare the power losses obtained with DIMAD and BDSIM along the ILC 
2~mrad and 20~mrad post-collision lines. A more comprehensive simulation 
study of the backgrounds from secondary particles will then follow.

\section*{Acknowledgement}
This work is supported by the Commission of the European Communities
under the 6$^{th}$ Framework Programme "Structuring the European
Research Area", contract number RIDS-011899. We would also like to 
thank Ilya Agapov, Grahame Blair and John Carter for the useful 
discussions and their assistance regarding the development of BDSIM.

\end{document}